\documentclass[prl,twocolumn,nofootinbib,reprint]{revtex4-1}
\usepackage{amsmath,amssymb,amsthm,bm,braket,ascmac,bbm}
\usepackage{graphicx}
\usepackage{color} 
\usepackage{hyperref}
\usepackage{cancel}

\theoremstyle{definition}

\begin{document}
\title{Axion Quality from Superconformal Dynamics
\\
}
\author{ Yuichiro Nakai and Motoo Suzuki}
\affiliation{\vspace{2mm} \\
Tsung-Dao Lee Institute and School of Physics and Astronomy, \\Shanghai
Jiao Tong University, 800 Dongchuan Road, Shanghai, 200240 China
}

\begin{abstract}

We discuss a possibility that a superconformal dynamics induces the emergence of a global $U(1)_{\rm PQ}$ symmetry
to solve the strong CP problem through the axion.
Fields spontaneously breaking the $U(1)_{\rm PQ}$ symmetry couple to new quarks
charged under the ordinary color $SU(3)_C$ and a new $SU(N)$ gauge group.
The theory flows into an IR fixed point where the $U(1)_{\rm PQ}$ breaking fields hold a large anomalous dimension
leading to the suppression of $U(1)_{\rm PQ}$-violating higher dimensional operators.
The spontaneous breaking of the $U(1)_{\rm PQ}$ makes the new quarks massive. 
The $U(1)_{\rm PQ}$ symmetry is anomalous under the $SU(3)_C$ but not under the $SU(N)$
so that the axion couples to only the color $SU(3)_C$ and the usual axion potential is generated.
We also comment on a model that the $U(1)_{\rm PQ}$ breaking fields are realized as meson superfields
in a new supersymmetric QCD.

 \end{abstract}

\begin{flushright}

\end{flushright}

\maketitle

{\bf Introduction.--}
The strong CP problem is an intriguing puzzle to motivate physics beyond the Standard Model (SM).
The current upper bound on the neutron electric dipole moment constrains the absolute value of the QCD vacuum angle $\bar{\theta}$
to be smaller than $10^{-11}$ \cite{Baker:2006ts,Afach:2015sja}.
Unlike other naturalness problems in the SM,
some shifts of $\bar{\theta}$ would not provide a visible change in our world.
The most common explanation for the strong CP problem is the introduction of a pseudo-Nambu-Goldstone boson,
called axion $a$ \cite{Weinberg:1977ma,Wilczek:1977pj},
associated with spontaneous breaking of a global $U(1)$ Peccei-Quinn ($U(1)_{\rm PQ}$) symmetry
\cite{Peccei:1977hh} (for reviews, see e.g. refs.~\cite{Kim:2008hd,DiLuzio:2020wdo}).
Non-perturbative QCD effects break the $U(1)_{\rm PQ}$ explicitly and
generate a potential of the axion whose minimum sets $\bar{\theta}$ to zero.
Astrophysical observations 
provide a lower limit on the $U(1)_{\rm PQ}$ breaking scale, $f_a \gtrsim 10^8 \, \rm GeV$
\cite{Chang:2018rso}.

A sufficiently small $\bar{\theta}$ requires the $U(1)_{\rm PQ}$ symmetry to be realized
to an extraordinary high degree. 
However, quantum gravity effects do not respect such a global symmetry.
We naturally expect $U(1)_{\rm PQ}$-violating higher dimensional operators
suppressed by appropriate powers of the Planck scale $M_{\rm Pl}$ \cite{Holman:1992us,Kamionkowski:1992mf,Barr:1992qq,Ghigna:1992iv,Carpenter:2009zs}.
Although a discrete $\mathbf{Z}_{n}$ symmetry can forbid some of the operators,
to suppress sufficiently higher order terms requires
$n \gtrsim 10$ 
which appears very contrived.
Other solutions to this axion quality problem have been explored by many authors.
They include
composite axion models
\cite{Kim:1984pt,Choi:1985cb,Randall:1992ut,Redi:2016esr,DiLuzio:2017tjx,Lillard:2017cwx,Lillard:2018fdt,Gavela:2018paw,Lee:2018yak},
models with a gauged symmetry ($e.g.$ $U(1)$) different from the $U(1)_{PQ}$
\cite{Cheng:2001ys,Harigaya:2013vja,Harigaya:2015soa,Fukuda:2017ylt,Fukuda:2018oco,Ibe:2018hir,Choi:2020vgb,Yin:2020dfn},
extra dimension models
\cite{Dienes:1999gw,Choi:2003wr,Flacke:2006ad,Cox:2019rro,Bonnefoy:2020llz,1842866}
and heavy axion models
\cite{Rubakov:1997vp,Berezhiani:2000gh,Hook:2014cda,Fukuda:2015ana,Gherghetta:2016fhp,Dimopoulos:2016lvn,Gherghetta:2020ofz}.

In this letter, we explore an alternative approach to the axion quality problem
that a superconformal dynamics induces the emergence of the $U(1)_{\rm PQ}$ symmetry.
Our model begins with the existence of a discrete $\mathbf{Z}_{N}$ with $N \sim 5$
which ensures that the model respects the $U(1)_{\rm PQ}$ symmetry at the renormalizable level.
We introduce a $SU(N)$ supersymmetric gauge theory with (anti-)fundamental quarks,
some of which are also charged under the ordinary color $SU(3)_C$.
The $\mathbf{Z}_{N}$ symmetry is anomaly-free under the $SU(3)_C$ as well as the $SU(N)$.
All the new quarks couple to fields responsible for the spontaneous $U(1)_{\rm PQ}$ breaking.
The theory flows into an IR fixed point where the $U(1)_{\rm PQ}$ breaking fields hold a large anomalous dimension.
Then, even if there exist higher dimensional operators dangerously violating the $U(1)_{\rm PQ}$ at the Planck scale,
those operators are significantly suppressed at low-energies.
The similar mechanism has been discussed in the context of the Nelson-Strassler model
to realize quark and lepton mass hierarchies
\cite{Nelson:2000sn} (for a more recent development using the $a$-maximization technique
\cite{Intriligator:2003jj}, see refs.~\cite{Poland:2009yb,Craig:2010ip}).
According to the AdS/CFT correspondence \cite{Maldacena:1997re}, the approach is similar to that of
the warped extra dimension model discussed in ref.~\cite{Cox:2019rro}.
However, to the best of our knowledge, our model is the first 4D calculable realization
to utilize a conformal dynamics to
suppress $U(1)_{\rm PQ}$-violating higher dimensional operators.
The spontaneous breaking of the $U(1)_{\rm PQ}$ makes all the new quarks massive.
The new quarks leading to a large anomalous dimension of the $U(1)_{\rm PQ}$ breaking fields also
play the role of the so-called KSVZ quarks
\cite{Kim:1979if,Shifman:1979if}.
Since the $U(1)_{\rm PQ}$ symmetry is anomalous under the $SU(3)_C$ but not under the $SU(N)$,
the axion couples to only the color $SU(3)_C$ and the usual axion potential is generated.
The $SU(N)$ finally confines and predicts the existence of $SU(N)$ glueballs.

While the $U(1)_{\rm PQ}$ breaking fields are introduced as elementary fields in the main part of the present work,
we will also comment on a possibility that they are realized as meson superfields in a new supersymmetric QCD (SQCD).
Interestingly, in the magnetic picture of the theory
\cite{Seiberg:1994pq,Intriligator:1995au},
the coupling of the $U(1)_{\rm PQ}$ breaking fields to dual quarks is automatic.

\begin{table}[ht]
\begin{center}
\begin{tabular}{|c||c|c|c|c|c|c|c|c|c|c|c|c|c|c|c|}
\hline
 & $Q_m$ &   $\bar Q_m$ & $Q_k$&   $\bar Q_k$ &  $\Phi$ &  $\bar\Phi$  \\ \hline
  $SU(N)$ & $\mathbf{N}$ & $\overline{\mathbf{N}}$ &$\mathbf{N}$ & $\overline{\mathbf{N}}$  & $\mathbf{1}$ & $\mathbf{1}$
\\ \hline
 $U(1)_{\rm PQ}~\left(\mathbf{Z}_{N}\right)$ & $+1$ & $0$ &$-1$ &$0$  & $-1$ & $+1$
\\ \hline
$U(1)_{R}$ & $\frac{N_f-N}{N_f}$ &$\frac{N_f-N}{N_f}$ &$\frac{N_f-N}{N_f}$ &$\frac{N_f-N}{N_f}$ & $\frac{2N}{N_f}$ & $\frac{2N}{N_f}$
\\ \hline
\end{tabular}
\end{center}
\caption{The charge assignments under the $SU(N)$ gauge group, the $U(1)_{\rm PQ}$ (and $\mathbf{Z}_{N}$)
and the anomaly-free $U(1)_R$
which determines anomalous dimensions of the fields.
Here, $m = 1, \cdots, N_f/2$ and $k = N_f/2 +1, \cdots, N_f$ where $N_f$ is even.
}
\label{tab:contents}
\end{table}%

{\bf  The model.--}
Let us consider a supersymmetric $SU(N)$ gauge theory
with $N_f$ vector-like pairs of chiral superfields in the (anti-)fundamental representation,
$Q_I, \bar Q_I$ $(I=1, \cdots, N_f)$.
Here, $N_f$ is assumed to be even.
We focus on $\frac{3}{2}N<N_f<3N$ where the theory is in conformal window
\cite{Intriligator:1995au}. 
To implement the QCD axion, we introduce two $SU(N)$ singlet chiral superfields $\Phi, \bar\Phi$
charged under the $U(1)_{\rm PQ}$ symmetry.
They are coupled to the new $SU(N)$ quarks in the superpotential,
\begin{align}
\label{KSVZ_potential}
W_Q=\lambda \Phi Q_m \bar Q_m+\bar\lambda \bar \Phi Q_k \bar Q_k\ ,
\end{align}
where $\lambda,~\bar\lambda$ denote dimensionless couplings,
$m$ runs from $1$ to $N_f/2$ and $k$ runs from $N_f/2 +1$ to $N_f$.
These terms explicitly break the original $SU(N_f)_L\times SU(N_f)_R$ flavor symmetries in the theory
into $SU(N_f/2)_1\times SU(N_f/2)_2$.
A subgroup $SU(3) \subset SU(N_f/2)_1$ is weakly gauged and regarded as the ordinary color $SU(3)_C$ in the SM.\footnote{
We can gauge a subgroup $SU(5) \subset SU(N_f/2)_1$ to accommodate the $SU(5)$ grand unified theory.
The following discussion is the same for this possibility.
}
Barring the effect of this $SU(3)_C$,
the couplings flow into $\lambda = \bar\lambda$ at low-energies.
The charge assignments under the $U(1)_{\rm PQ}$ symmetry are summarized in Tab.~\ref{tab:contents}.
The $U(1)_{\rm PQ}$ symmetry is not anomalous under the $SU(N)$ but has the $U(1)_{\rm PQ}-SU(3)_C-SU(3)_C$ anomaly
whose coefficient is given by $A_{U(1)_{\rm PQ}-SU(3)_C-SU(3)_C}=N$.
Then, 
an anomaly-free discrete symmetry $\mathbf{Z}_{N} \subset U(1)_{\rm PQ}$ is realized,
which leads to the $U(1)_{\rm PQ}$ symmetry at the renormalizable level.
Below, we will discuss Planck-scale suppressed $U(1)_{\rm PQ}$-violating operators,
but those operators must respect the $\mathbf{Z}_{N}$ symmetry.
The fields $\Phi, \bar\Phi$ obtain a non-zero vacuum expectation value (VEV) via the superpotential,
\begin{align}
\label{Xsuper}
    W'_X&=
     \kappa' X(2 \Phi \bar{\Phi}-f'^{2}) \, ,
\end{align}
which breaks the $U(1)_{\rm PQ}$ symmetry spontaneously.
Here, $X$ is a singlet chiral superfield, $\kappa'$ is a dimensionless parameter and
$f'$ is a constant with a mass dimension.\footnote{The superpotential $W'_X$ explicitly breaks
the anomaly-free $U(1)_R$ symmetry in the gauge theory.
We assume that $\kappa'$ does not enter a fixed point.}

The gauge theory is in conformal window and believed to have a non-trivial IR fixed point. 
Here, let us assume the $SU(N)$ gauge coupling $g$, $\lambda$ and $\bar\lambda$ approach values at the fixed point
and the theory is in the conformal regime between the energy scales $\Lambda$ and $M_c$ ($\Lambda>M_c$). 
We will demonstrate the existence of the IR fixed point later.
In this regime, the conformal dynamics generates a large anomalous dimension of $\Phi, \bar\Phi$
through the superpotential terms of Eq.~\eqref{KSVZ_potential}.
The wave function renormalization factor of $\Phi$ (and $\bar\Phi$) at IR is given by
\begin{align}
    Z_\Phi=\left(\frac{M_c}{\Lambda}\right)^{-\gamma_\Phi}\ ,
    \label{wavefunction}
\end{align}
where $\gamma_{\Phi} = 6\frac{N}{N_f}-2$ is the anomalous dimension of $\Phi$
which is exactly determined in terms of the anomaly-free $U(1)_R$ charges summarized in Tab.~\ref{tab:contents}.
We now canonically normalize $\Phi$ as
\begin{align}
  \Phi= \left(\frac{M_c}{\Lambda}\right)^{\gamma_\Phi/2}\hat\Phi\ ,
\end{align}
whose hat $\hat~$ denotes a field in the canonical normalization.
Then, the superpotential \eqref{Xsuper} is rewritten in terms of the normalized fields,
\begin{align}
    W_X&=
    \kappa \left(\frac{M_c}{\Lambda}\right)^{\gamma_\Phi} X(2\hat{\Phi}\hat{\bar\Phi}-f^{2}) \, ,
\end{align}
where $\kappa \sim \kappa'$ is dimensionless and
$f\sim\left(\frac{M_c}{\Lambda}\right)^{-\gamma_\Phi/2}f'$ is a constant with a mass dimension.
The $U(1)_{\rm PQ}$ breaking scale is determined by $f$ which also gives the conformal breaking, $M_c \sim f$.
The wave function renormalization factor of Eq.~\eqref{wavefunction} will play a key role in suppressing
$U(1)_{\rm PQ}$-violating higher dimensional operators as we will see below.

Once the $U(1)_{\rm PQ}$ breaking fields $\Phi, \bar\Phi$ obtain the VEV,
all the new quarks $Q_I, \bar Q_I$ become massive, and then the axion-gluon coupling is generated
in the effective Lagrangian after the integration of the new quarks,
\begin{align}
\label{eq:axion_gluon_coupling}
   \mathcal{L}_{\rm eff} \supset N\frac{a}{F_a}\frac{g_c^2}{32\pi^2}G\tilde G\ ,
\end{align}
where $a$ denotes the axion, $G$ is the field strength of the gluon, $\tilde G$ is its dual, $g_c$ is the QCD gauge coupling constant and $F_a /N =\sqrt{2}f /N$ is the axion decay constant.
The same axion-gluon coupling is obtained in the KSVZ axion model~\cite{Kim:1979if,Shifman:1979if}
with $N$ flavors of $SU(3)_C$ vector-like quarks.
Since the $U(1)_{\rm PQ}$ symmetry is not anomalous under the $SU(N)$,
the terms in Eq.\,\eqref{KSVZ_potential} do not lead to the axion-$SU(N)$ gluon coupling even after the integration of the quarks. 
The axion potential is obtained via the non-perturbative QCD effect,
\begin{align}
\label{QCDpotential}
V\sim  m_{\pi}^2f_{\pi}^2\cos\left(N\frac{a}{F_a}\right)\ ,
\end{align}
where $m_{\pi}$ and $f_{\pi}$ are the pion mass and the decay constant respectively 
and $m_{\pi}^2f_{\pi}^2=(0.1\,{\rm GeV})^4$.
Then, the strong CP problem is solved in the ordinary way.
After the decoupling of $Q_I, \bar Q_I$,
the model becomes a $SU(N)$ pure Yang Mills theory.
Because of a large gauge coupling of the $SU(N)$ at the fixed point,
the theory confines just below the conformal breaking scale $M_c$
and predicts heavy $SU(N)$ glueballs and their superpartners.

{\bf Axion quality.--}
To address the axion quality problem, explicit $U(1)_{\rm PQ}$ breaking terms must be highly suppressed
compared to the axion potential generated by the non-perturbative QCD effect \eqref{QCDpotential}.
The most dangerous Planck-scale suppressed operator respecting the $\mathbf{Z}_{N}$ symmetry
is the superpotential term such as
\begin{align}
\label{eq:PQ_breaking_op_1}
    W_{\cancel{\rm PQ}}&\sim \frac{\Phi^{N}}{M_{\rm Pl}^{N-3}}
    \sim  \left(\frac{M_c}{\Lambda}\right)^{\frac{N\gamma_{\Phi}}{2}}\frac{\hat{\Phi}^{N}}{M_{\rm Pl}^{N-3}}
     \ ,
\end{align}
which leads to the scalar potential in supergravity with $e.g.$ the constant term $W=m_{3/2}M_{\rm Pl}^2$ of the superpotential via $V\supset-3W W^*/M_{\rm Pl}^2$,
\begin{align}
\label{eq:PQ_breaking_scalar_1}
    V_{\cancel{\rm PQ}}= \left(\frac{M_c}{\Lambda}\right)^{\frac{N\gamma_{\Phi}}{2}}\frac{\kappa_{\cancel{\rm PQ}}\, m_{3/2}\hat{\Phi}^{N}}{M_{\rm Pl}^{N-3}}\ .
\end{align}
Here, $m_{3/2}$ is the gravitino mass, $\kappa_{\cancel{\rm PQ}}$ is a model dependent coefficient, and
$\hat\Phi$ denotes the scalar component which is the same notation as the chiral superfield for notational simplicity.
The $U(1)_{\rm PQ}$-violating axion potential is then obtained as
\begin{align}
\label{PQVpotential}
    V_{\cancel{\rm PQ}} \supset  \left(\frac{M_c}{\Lambda}\right)^{N(3N/N_f-1)}\frac{\kappa_{\cancel{\rm PQ}}\, m_{3/2} F_a^{N}}{M_{\rm Pl}^{N-3}} \cos\left(N\frac{a}{F_a}+\varphi\right) ,
\end{align}
where $\varphi$ denotes a CP phase and $\gamma_{\Phi}=6\frac{N}{N_f}-2$ has been used.
We now define the axion quality factor $\mathcal{Q}$ by
\begin{align}
V_{\cancel{\rm PQ}}\equiv \mathcal{Q}\, m_{\pi}^2f_{\pi}^2\cos\left(N\frac{a}{F_a}+\varphi\right) .
\end{align}
Assuming $\varphi=\mathcal{O}(1)$,
the experimental upper bound on the $\bar \theta$ parameter
\cite{Baker:2006ts,Afach:2015sja} requires $\mathcal{Q}\lesssim 10^{-10}$
to secure the axion quality.

\begin{figure}
\begin{minipage}[t]{\hsize}
\includegraphics[width=8cm]{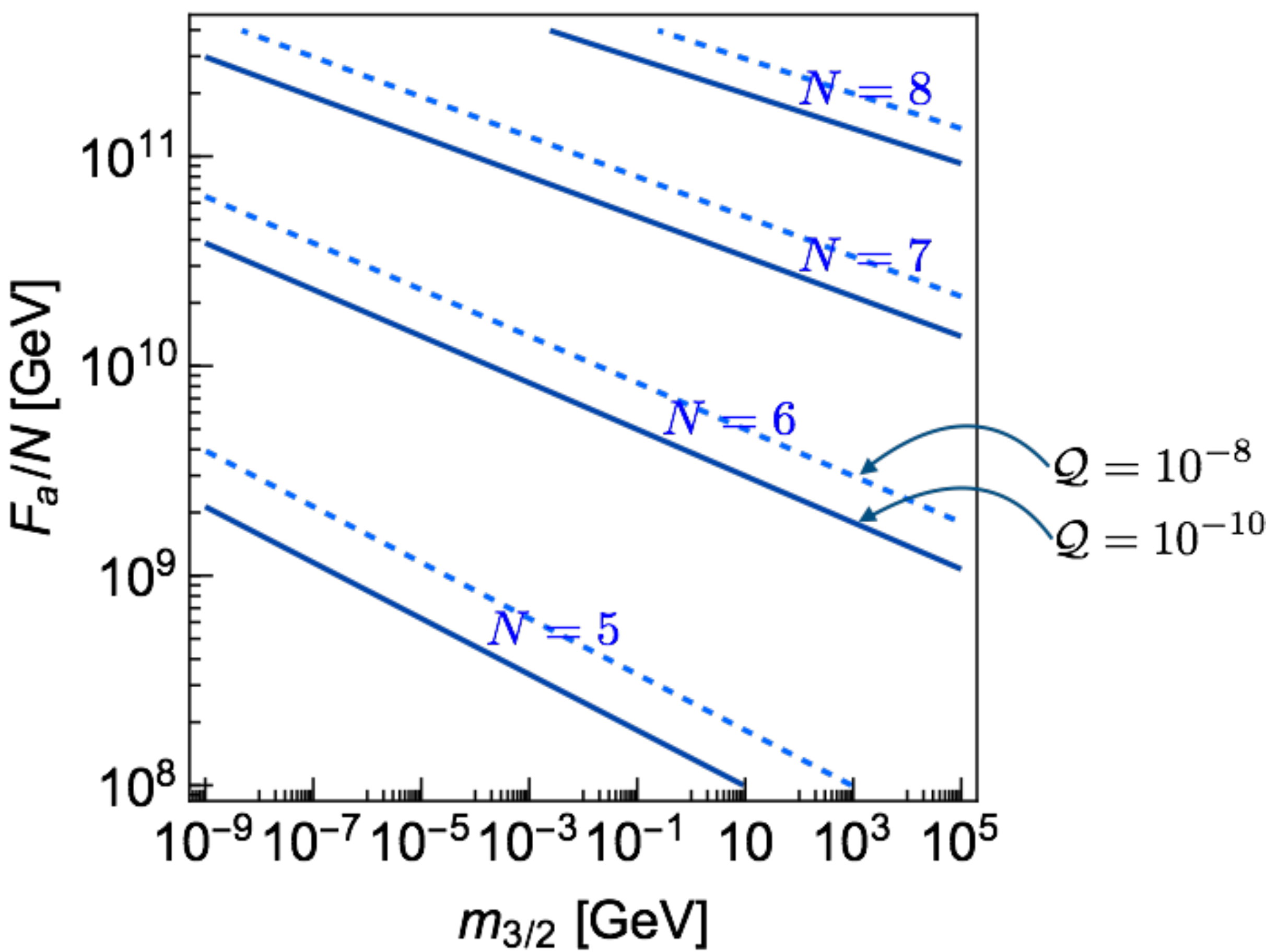}
  \end{minipage}
  \caption{The contours of the quality factor $\mathcal{Q}$
  calculated from the potential \eqref{PQVpotential} in the $m_{3/2}-F_a/N$ plane.
  We take $N_f=2\,N$,~$M_c=F_a$,~$\kappa_{\cancel{PQ}}=1$ and $\Lambda=0.1\,M_{\rm Pl}$.
  The solid and dashed lines correspond to the quality factor $\mathcal{Q}=10^{-10},~10^{-8}$, respectively.
}
\label{fig:quality_factor} 
\end{figure}

Fig.~\ref{fig:quality_factor} shows the contours of $\mathcal{Q}$ calculated from the potential \eqref{PQVpotential}
in the $m_{3/2}-F_a/N$ plane.
Here, we take $N_f=2\,N$,~$M_c=F_a$,~$\kappa_{\cancel{PQ}}=1$, and $\Lambda=0.1\,M_{\rm Pl}$.
The solid and dashed lines denote the quality factor $\mathcal{Q}=10^{-10},~10^{-8}$, respectively.
The axion decay constant $F_a/N$ is constrained from the supernova 1981A observation, $F_a/N\gtrsim 10^8\,{\rm GeV}$~\cite{Chang:2018rso}.
We can see from the figure that there is a parameter space to solve the axion quality problem for $N \geq 5$.
While the case of $N=4$ is not shown in the figure,
$\mathcal{Q}=10^{-5}$ is obtained for $F_a/N\sim 10^8$\,GeV and $m_{3/2}\sim 1$\,eV.

Other potentially dangerous $U(1)_{\rm PQ}$-violating operators are
\begin{equation}
\begin{split}
\label{eq:PQ_breaking_op_2}
W'_{\cancel{\rm PQ}} &\sim \frac{({Q}_m {\bar Q}_m)^{N-k} {\bar\Phi}^k}{M_{\rm Pl}^{2N-k-3}} \\[1ex]
&\sim \frac{(\hat{Q}_m\hat{\bar Q}_m)^{N-k}\hat{\bar\Phi}^k}{M_{\rm Pl}^{2N-k-3}}\left(\frac{\Lambda}{M_c}\right)^{\gamma_\Phi\frac{N-2k}{2}},
\end{split}
\end{equation}
with $k=0, \cdots, N-1$.
While these operators will not lead to the axion potential by themselves,
we must be careful because they are enhanced at low-energies due to the negative anomalous dimension of $Q\bar Q$.
However, for $e.g.$ $N_f=2N$, $\Phi~(\bar\Phi)$ and $Q\bar Q$ have the same scaling dimension $3/2$, and then Eq.\,\eqref{eq:PQ_breaking_op_2} can be rewritten as
\begin{equation}
\begin{split}
W'_{\cancel{\rm PQ}}&\sim \left(\frac{M_c}{\Lambda}\right)^{\frac{N}{2}} \!\! \frac{\hat{\bar\Phi}^N}{M_{\rm Pl}^{N-3}}
\left(\frac{\hat{Q}_m\hat{\bar Q}_m}{\hat{\bar\Phi}\,M_c}\right)^{N-k} \! \left(\frac{\Lambda}{M_{\rm Pl}}\right)^{N-k},
\end{split}
\end{equation}
which is suppressed compared to Eq.\,\eqref{eq:PQ_breaking_op_1} for $\langle\hat{\bar\Phi} \rangle\approx M_c$ and $\Lambda<M_{\rm Pl}$.
We also note that $U(1)_{\rm PQ}$-violating operators in the K{\"a}hler potential are negligible compared to
those in the superpotential.

\begin{figure*}[ht]
\includegraphics[width=7.5cm]{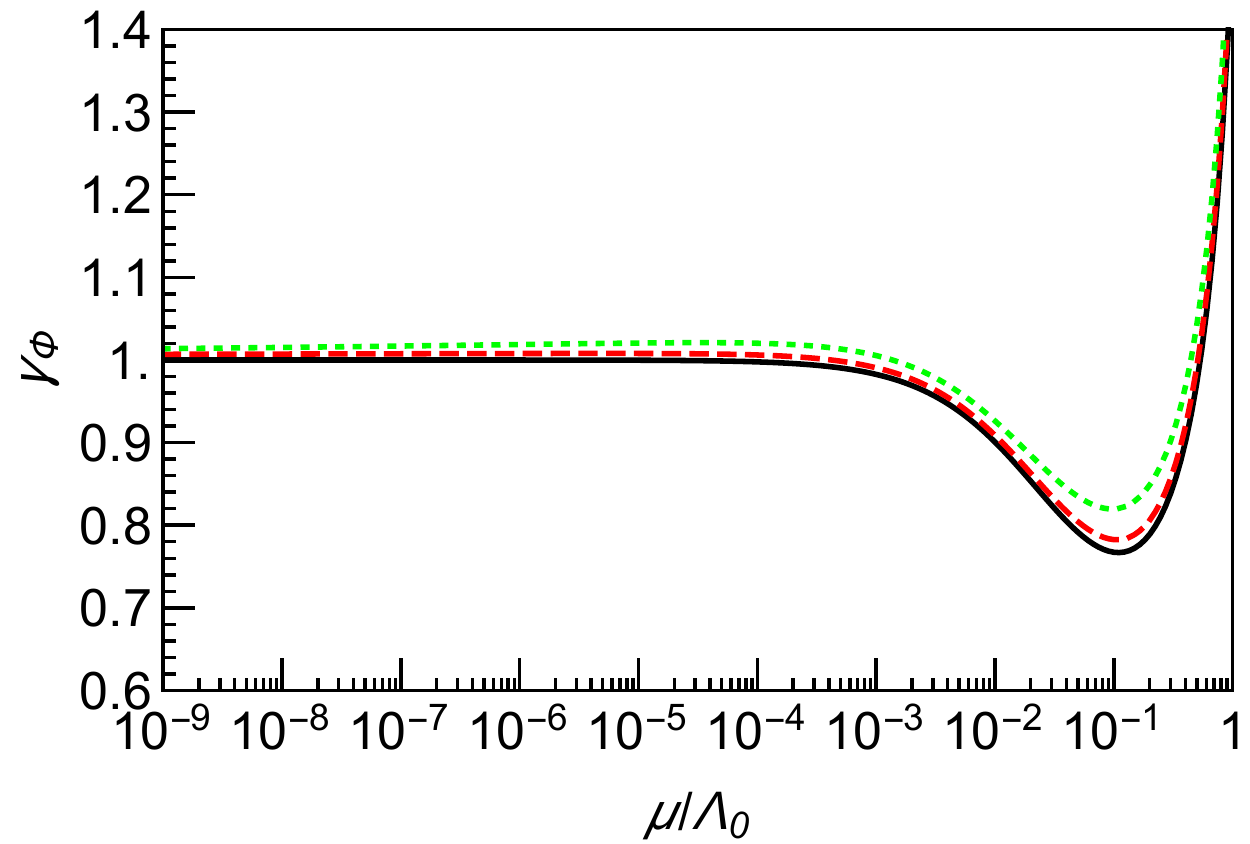}~~~~~~~~~~
\includegraphics[width=7.5cm]{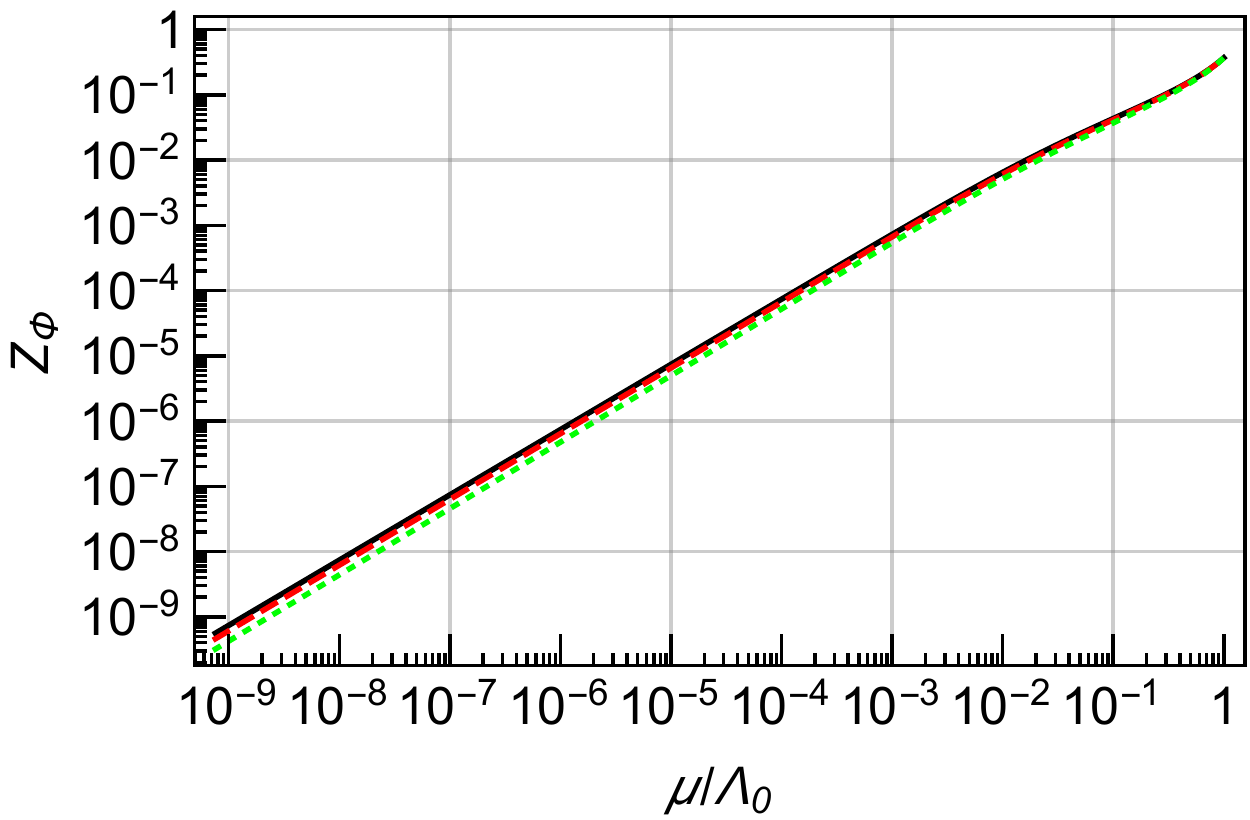}
  \caption{
  {\it Left panel} : The flow of $\gamma^2_\Phi$ for $g_c=0,1,2$ and $g=\lambda_1=\lambda_2=\bar\lambda=2$
  at $\Lambda_0$
  denoted by the black solid, red dashed, green dotted lines, respectively.
  {\it Right panel} : The flow of the wave function renormalization factor $Z_\Phi$
  for $g_c=0,1,2$ and $g=\lambda_1=\lambda_2=\bar\lambda=2$ at $\Lambda_0$.
  The color code and the line style are the same as those of the left panel.
}
\label{fig:z_phi} 
\end{figure*}

{\bf The IR fixed point.--}
Let us now discuss the existence of the IR fixed point for the $SU(N)$ gauge coupling $g$ and $\lambda, \bar\lambda$
in the superpotential \eqref{KSVZ_potential}.
We first ignore the effect of the $SU(3)_C$ gauge coupling
and solve the renormalization group equations (RGEs) for $g$, $\lambda$ and $\bar\lambda$,
\begin{equation}
\begin{split}
    \frac{dg}{dt}&=-\frac{g^3}{2}\frac{b_0+\frac{1}{2}\sum_{m}\left(\gamma^1_{Q_m}+\gamma^1_{\bar Q_m}\right)+\frac{1}{2}\sum_{k}\left(\gamma^1_{Q_k}+\gamma^1_{\bar Q_k}\right)}{8\pi^2-C_A g^2} , \\[1ex]
    \frac{d\lambda}{dt}&=\frac{\lambda}{2}\left(\gamma^1_\Phi+
    \gamma^1_{Q_m}+\gamma^1_{\bar Q_m}
    \right) , \\[1ex]
    \frac{d\bar\lambda}{dt}&=\frac{\bar\lambda}{2}\left(\gamma^1_{\bar\Phi}+
    \gamma^1_{Q_k}+\gamma^1_{\bar Q_k}
    \right) ,
\end{split}
\end{equation}
where $t=\ln(\mu/\Lambda_0)$ with $\mu$ being the RG scale, $C_A=N$ and $b_0=3N-N_f$.
Here, we use the exact NSVZ $\beta$ function \cite{Novikov:1983uc,Novikov:1985ic,Novikov:1985rd}
for the RGE of the gauge coupling,
while the RGEs of $\lambda$ and $\bar\lambda$ are shown at the one-loop level.
The anomalous dimensions are given by 
\begin{equation}
\begin{split}
\label{eq:gamma_Q_m}
    &\gamma^1_{Q_m}=\gamma^1_{\bar Q_m}=-\frac{1}{8\pi^2}\left(2C_2g^2 -\lambda^2\right), \\
    &\gamma^1_{Q_k}=\gamma^1_{\bar Q_k}=-\frac{1}{8\pi^2}\left(2C_2g^2 -\bar\lambda^2\right) , \\
    &\gamma^1_\Phi=\frac{1}{8\pi^2}\lambda^2N N_f/2\ , \\
    &\gamma^1_{\bar\Phi}=\frac{1}{8\pi^2}\bar\lambda^2N N_f/2 \ ,
\end{split}
\end{equation}
with $C_2=\frac{N^2-1}{2N}$.
We also calculate the RGEs for $\lambda$ and $\bar\lambda$ at the two-loop level
whose expressions are summarized in appendix.
Fig.~\ref{fig:flow} shows the RG flows of $g$ and $\lambda$ from a scale $\Lambda_0$
to $\mu=10^{-9}\Lambda_0$ for different initial values as a demonstration.
We take $N=5$, $N_f=10$ and $\bar\lambda=2$ at $\Lambda_0$.
Blue and red dots correspond to the cases using the one and two-loop RGEs for $\lambda$, respectively.
The figure illustrates both couplings flow into a non-trivial IR fixed point.
The blue circle around the center denotes the values of $g$ and $\lambda$
obtained by comparing the anomalous dimensions at one-loop \eqref{eq:gamma_Q_m} to
those determined by the $U(1)_R$ charges in Tab.~\ref{tab:contents},
\begin{equation}
\begin{split}
\label{eq:fixed_point}
   &\frac{g_*^2}{8\pi^2} \simeq\frac{N}{N^2-1}\frac{\gamma_\Phi}{2}\left(1+\frac{2}{N N_f/2}\right) , \\
   &\frac{\lambda_*^2}{8\pi^2} \simeq \frac{\gamma_\Phi}{(N_f/2) N}\ ,
\end{split}
\end{equation}
where $\gamma_{\Phi} = 6\frac{N}{N_f}-2$.
The anomalous dimensions up to the two-loop order \eqref{eq:gamma_Q_m_2} are used to find the values of the couplings
at the red circle.
We also plot the anomalous dimension of $\Phi$ at two-loop $\gamma^2_\Phi$ in the left panel of Fig.~\ref{fig:z_phi}
(black solid). 
We take $N=5$, $N_f=10$ and $g=\lambda=\bar\lambda=2$ at the initial scale $\Lambda_0$.
The figure indicates that $\gamma^2_\Phi$ converges to $\gamma_\Phi = 6\frac{N}{N_f}-2 =1$.
Therefore, the theory is expected to enter the conformal regime in the IR region as we have assumed in the above discussion.

\begin{figure}[!t]
\hspace{-0.5cm}
\includegraphics[width=7.5cm]{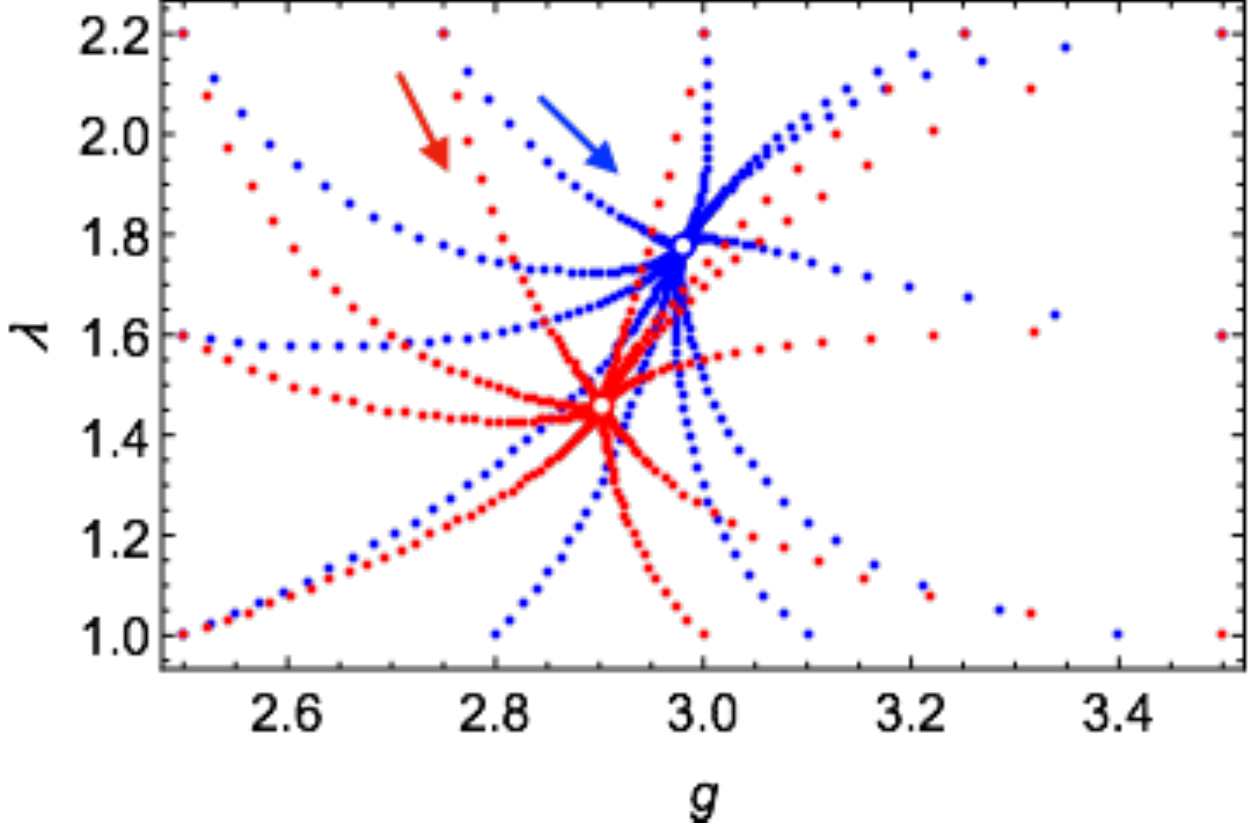}
  \caption{The RG flows of $g$ and $\lambda$ from a scale $\Lambda_0$
  to $\mu=10^{-9}\Lambda_0$ for different initial values.
  Blue and red dots correspond to the cases using the one and two-loop RGEs for $\lambda$, respectively.
  The arrows show the directions of the flows.
  We take $N=5$, $N_f=10$ and $\bar\lambda=2$ at $\Lambda_0$.
  The blue circle around the center denotes the IR fixed point of Eq.\,\eqref{eq:fixed_point}
  which is obtained by comparing the anomalous dimensions at one-loop \eqref{eq:gamma_Q_m} to
  those determined by the $U(1)_R$ charges.
  The anomalous dimensions up to the two-loop order \eqref{eq:gamma_Q_m_2} are used to find the values of the couplings
  at the red circle.
}
\label{fig:flow} 
\end{figure}

The IR fixed point can be disturbed by the $SU(3)_C$ gauge coupling.
To discuss this effect, we first decompose the superpotential term in Eq.~\eqref{KSVZ_potential} as
\begin{align}
    W_Q \supset \lambda\, Q_m\bar Q_m \to \lambda_1\,\Phi Q_a \bar Q_a+ \lambda_2\,\Phi Q_\alpha\bar Q_\alpha\ ,
\end{align}
where $Q_a, \bar Q_a$ $(a=1,2,3)$ denote the fundamental and anti-fundamental representations of the $SU(3)_C$ gauge group,
$Q_\alpha, \bar Q_\alpha$ $(\alpha=4, \cdots , N_f/2)$ are the quarks that are not charged under the $SU(3)_C$
and $\lambda_{1,2}$ are dimensionless couplings.
The anomalous dimensions including the $SU(3)_C$ effect at the two-loop level are summarized in appendix. 
We use the one-loop RGE for the $SU(3)_C$ gauge coupling, 
\begin{align}
    \frac{dg_c}{dt}=-\frac{1}{16\pi^2}g_c^3 b_3\ ,
\end{align}
which is solved as
\begin{align}
    \frac{4\pi}{g_c^2}=\left.\frac{4\pi}{g_c^2}\right|_{\mu=M_c}+\frac{b_3}{2\pi}\ln(\mu/M_c)\ ,
\end{align}
where we take $b_3=3-N$ for $\mu>M_c$ by assuming all the new quarks have masses around $M_c$.
Here, the factor $3$ is from the MSSM particles and the factor $-N$ is from the $Q_a, \bar Q_a$ quarks.
For $N=5$, the $SU(3)_C$ gauge coupling becomes asymptotic non-free.
In this case, we obtain $g_c\approx 1$ around $\mu = 10^{17} \, \rm GeV$
for the spectrum of the MSSM particles at about $10 \, \rm TeV$ and $4\pi/g_c^2|_{\mu=M_c}\approx 20$
at $M_c = 10^8 \, \rm GeV$.
We numerically solve the two-loop RGEs from a scale $\Lambda_0$
to $\mu=10^{-9}\Lambda_0$. 
The left panel of Fig.~\ref{fig:z_phi} shows the flow of $\gamma^2_\Phi$ for $g_c=1,2$ at $\Lambda_0$
denoted by the red dashed and green dotted lines, respectively.
The initial values of the couplings at $\Lambda_0$ are $g=\lambda_1=\lambda_2=\bar\lambda=2$.
We also plot the flow of the wave function renormalization factor $Z_\Phi$ for $g_c=0,1,2,3$
and $g=\lambda_1=\lambda_2=\bar\lambda=2$ at $\Lambda_0$
in the right panel of Fig.~\ref{fig:z_phi}.
From the figures, we can confirm that $\gamma^2_\Phi$ converges into the one without the $SU(3)_C$ effect
and the smallness of $Z_\Phi$ enables to solve the axion quality problem.


{\bf A model with the dual picture.--}
So far, we have discussed the model where the $U(1)_{\rm PQ}$ breaking fields are introduced as elementary fields,
but here let us comment on a possibility that they are realized as meson superfields in a new SQCD.
Consider a $SU(N_f - N)$ SQCD with $N_f$ vector-like pairs of quarks
whose dual magnetic picture is given by a $SU(N)$ SQCD with the same number of flavors $D_i, \bar D^i$ $(i=1, \cdots ,N_f)$
\cite{Seiberg:1994pq}.
In the magnetic theory, there also exist meson chiral superfields $\mathcal{M}^i_j$ which are coupled to the dual quarks 
through the superpotential,
\begin{align}
\label{eq:magnetic_superpotential}
   W_{\rm mag}=y\,\mathcal{M}^i_j D_i \bar D^j\ ,
\end{align}
where $y$ is a dimensionless coupling.
For $\frac{3}{2}N\leq N_f \leq 3N$, this gauge theory is in conformal window for both the electric and magnetic pictures
and flows into an IR fixed point.
We now gauge a diagonal $SU(3)$ subgroup of the $SU(N_f)_L\times SU(N_f)_R$ flavor symmetry in the theory
and identify it as the SM color gauge group.
For notational convenience,  
we decompose the mesons $\mathcal{M}^i_j$ into
\begin{align}
\label{meson}
    \mathcal{M}^i_j=\left(
    \begin{array}{ccc}
    {\mathcal{M}}^{a}_{1b}
    & {\mathcal{M}}^{a}_{4\beta} & {\mathcal{M}}^{a}_{6\bar j}\\
    {\mathcal{M}}^{\alpha}_{5b} &
    {\mathcal{M}}^{\alpha}_{2\beta}
    & {\mathcal{M}}^{\alpha}_{8\bar j}\\
    {\mathcal{M}}^{\bar i}_{7b} & {\mathcal{M}}^{\bar i}_{9\beta} & 
    {\mathcal{M}}^{\bar i}_{3\bar j}
    \end{array}
    \right) ,
\end{align}
where $a, b$ ($=1,2,3$) denote the color $SU(3)_C$ indices, $\alpha, \beta =4,5,6$
and $\bar i, \bar j =7, \cdots ,N_f$. 
The $U(1)_{\rm PQ}$ charges are, for example, assigned as shown in Tab.~\ref{tab:contents_magnetic}. 
With these assignments, the $U(1)_{\rm PQ}$ symmetry is not anomalous under the $SU(N)$
but is anomalous under the $SU(3)_C$. 
With the decomposition of Eq.~\eqref{meson}, we can see that the superpotential \eqref{eq:magnetic_superpotential} contains
the terms similar to those introduced in Eq.~\eqref{KSVZ_potential},
\begin{align}
\label{eq:KSVZ_coupling}
    W_{\rm mag} \supset y\, \mathcal{M}_1 D_a \bar D^a
    + y\,\mathcal{M}_2 D_{\alpha}\bar D^{\alpha}\ .
\end{align}
Here,
we have defined $\mathcal{M}_{1} \equiv \frac{1}{3}\mathcal{M}_{1a}^a$
and $\mathcal{M}_{2} \equiv \frac{1}{3}\mathcal{M}_{2\alpha}^{\alpha}$. 
Note that $\mathcal{M}_{1,2}$ are color singlet but $U(1)_{\rm PQ}$ charged.
Once they obtain non-zero VEVs, we get the axion-gluon coupling \eqref{eq:axion_gluon_coupling}.
As before, the $U(1)_{\rm PQ}$ symmetry at the renormalizable level
is ensured by an anomaly-free $\mathbf{Z}_{N} \subset U(1)_{\rm PQ}$. 
Explicit $U(1)_{\rm PQ}$-violating higher dimensional operators are suppressed
due to large anomalous dimensions of $\mathcal{M}_{1,2}$.

Several comments are in order.
The IR fixed point can be disturbed by the $SU(3)_C$ gauge interaction.
In order to keep the electric/magnetic duality reliable, 
the values of the couplings in both electric and magnetic pictures at the fixed point
must be much larger than the QCD gauge coupling,
which requires the theory to be near the middle of conformal window, $N_f\approx 2\,N$.
Extra meson and quark chiral superfields must get masses appropriately.
In particular, $SU(3)_C$-charged mesons must be stabilized at the origin to avoid the color breaking.
If $\mathcal{M}_{3} \equiv \frac{1}{N_f-6}\mathcal{M}_{3\bar i}^{\bar i}$ obtains a non-zero VEV,
all the quarks become massive.
Below the scales of $\mathcal{M}_{1,2,3}$ VEVs, the model becomes a confining $SU(N)$ pure Yang Mills theory.
Further explorations of this model are left to a future study.

\begin{table*}[ht]
\begin{center}
\begin{tabular}{|c||c|c|c|c|c|c|c|c|c|c|c|c|c|c|c|}
\hline
 & $D_a$&  $\bar D^a$&  $D_{\alpha}$&   $\bar D^{\alpha}$ &   $D_{\bar i}$ & $\bar D^{\bar i}$ &  $\mathcal{M}_1$ &  $\mathcal{M}_2$ &  $\mathcal{M}_3$&  $\mathcal{M}_4$ &  $\mathcal{M}_5$ &  $\mathcal{M}_6$ &  $\mathcal{M}_7$ &  $\mathcal{M}_8$  &  $\mathcal{M}_9$ \\ \hline
 $SU(3)_C$ & $\bar{\mathbf{3}}$ &$\mathbf{3}$ &$\mathbf{1}$ &$\mathbf{1}$ &$\mathbf{1}$  & $\mathbf{1}$ & $\mathbf{Adj. + 1}$ & $\mathbf{1}$& $\mathbf{1}$& $\mathbf{3}$ & $\bar{\mathbf{3}}$ & $\mathbf{3}$ & $\bar{\mathbf{3}}$ & $\mathbf{1}$ & $\mathbf{1}$
\\ \hline
 $U(1)_{\rm PQ}~\left(\mathbf{Z}_{N}\right)$ & $+1$ &$0$ &$0$ &$-1$ &$0$  & $0$ & $-1$ & $+1$& $0$& $0$ & $0$ & $-1$ & $0$ & $0$ & $+1$
\\ \hline
\end{tabular}
\end{center}
\caption{The matter content of the magnetic picture of the model and the charge assignments under
the color $SU(3)_C$ and the $U(1)_{\rm PQ}$ (and $\mathbf{Z}_{N}$).
Here, $a$ ($=1,2,3$) denotes the color $SU(3)_C$ index, $\alpha =4,5,6$
and $\bar i  =7, \cdots ,N_f$. 
}
\label{tab:contents_magnetic}
\end{table*}%

{\bf Conclusions and discussions.--}
We have considered a possibility that a superconformal dynamics helps to solve the strong CP problem
through the axion with a sufficient quality.
The $U(1)_{\rm PQ}$ breaking fields are coupled to the new quarks charged
under the $SU(3)_C$ and the new $SU(N)$.
The theory flows into a non-trivial IR fixed point
where the $U(1)_{\rm PQ}$ breaking fields hold a large anomalous dimension leading to a strong suppression of
explicit $U(1)_{\rm PQ}$ breaking operators.
The $U(1)_{\rm PQ}$ is anomalous under the $SU(3)_C$ but not under the $SU(N)$
so that the usual axion potential is generated by non-perturbative QCD effects.

The model respects the anomaly-free $\mathbf{Z}_{N} \subset U(1)_{\rm PQ}$, which realizes the $U(1)_{\rm PQ}$ symmetry at the renormalizable level.
If the $U(1)_{\rm PQ}$ is spontaneously broken after the end of inflation,
cosmic strings are formed at a temperature close to the $U(1)_{\rm PQ}$ breaking scale
(see $e.g.$ ref.~\cite{Kawasaki:2013ae} for a review on axion cosmology).
Below around the QCD temperature, domain walls attached to the cosmic strings are formed.
They are stable due to the $\mathbf{Z}_{N}$ symmetry and cause a cosmological problem.
In order to avoid this, the $U(1)_{\rm PQ}$ symmetry must be broken before the end of inflation.
In this case, the axion isocurvature perturbation is produced,
which leads to a constraint on the Hubble scale of inflation, $H_{\rm inf}\lesssim 10^7$\,GeV.
Cosmological aspects might be an interesting future direction.

We may be able to use the same superconformal dynamics to realize the quark and lepton mass hierarchies in the same way as
the Nelson-Strassler model \cite{Nelson:2000sn}.
Such a possibility has been recently discussed in the 5D context
\cite{Bonnefoy:2020llz}.
One extra benefit of this scenario is that flavor-dependent soft scalar masses are automatically suppressed
\cite{Kobayashi:2001kz,Nelson:2001mq} (see also ref.~\cite{Kobayashi:2010ye}).

%
\section*{Acknowledgements}

We would like to thank Ryosuke Sato for discussions
and helpful comments on the manuscript.
We are also grateful to Kavli IPMU for their hospitality during the COVID-19 pandemic.

%

\vspace{0.5cm}

{\bf Appendix: the two-loop RGEs.--}
Here, we summarize the expressions of the two-loop RGEs for $g$, $\lambda_1$, $\lambda_2$ and $\bar\lambda$.
The effect of the $SU(3)_C$ gauge coupling is included.
They are given by
\begin{equation}
\begin{split}
    &\frac{dg}{dt}=-\frac{g^3}{2}\frac{b_0+\frac{1}{2}\sum_{i=a,\alpha,k}
    \left(\gamma^2_{Q_i}+\gamma^2_{\bar Q_i}\right)}{8\pi^2-C_A g^2} , \\[1ex]
    &\frac{d\lambda_1}{dt}=\frac{\lambda_1}{2}\left(\gamma^2_\Phi+
    \gamma^2_{Q_a}+\gamma^2_{\bar Q_a}
    \right) , \\[1ex]
    &\frac{d\lambda_2}{dt}=\frac{\lambda_2}{2}\left(\gamma^2_\Phi+
    \gamma^2_{Q_\alpha}+\gamma^2_{\bar Q_\alpha}
    \right) , \\[1ex]
    &\frac{d\bar\lambda}{dt}=\frac{\bar\lambda}{2}\left(\gamma^2_{\bar\Phi}+
    \gamma^2_{Q_k}+\gamma^2_{\bar Q_k}
    \right) ,
\end{split}
\end{equation}
with the anomalous dimensions at the two-loop level,
\begin{equation}
\begin{split}
    \gamma^2_{Q_a}&=\gamma^2_{\bar Q_a}\\
    &=-\frac{1}{8\pi^2}\left(2C_2g^2 +2C_2'g_c^2-\lambda_1^2\right)\\
    &+\frac{2}{(16\pi^2)^2}\left[-\lambda_1^4-3N\lambda_1^4-N(N_f/2-3)\lambda_1^2\lambda_2^2\right.\\
    &\left.+2g^4(C_2 S_N(R)+2C_2^2-3C_N(G)C_2)\right.\\
    &\left.+2g_c^4(C_2'S_3(R)+2{C_2'}^2-3C_3(G)C_2')\right.\\
    &\left.+8g_c^2 g^2 C_2 C_2' 
    \right],\\
    \gamma^2_{Q_\alpha}&=\gamma^2_{\bar Q_\alpha}\\
    &=-\frac{1}{8\pi^2}\left(2C_2g^2-\lambda_2^2\right)\\
    &+\frac{2}{(16\pi^2)^2}\left[-\lambda_2^4-(N_f/2-3)N\lambda_2^4-3N\lambda_1^2\lambda_2^2\right.\\
    &\left.+2g^4(C_2S_N(R)+2C_2^2-3C_N(G)C_2)
    \right],\\
    \gamma^2_{Q_k}&=\gamma^2_{\bar Q_k}\\
    &=-\frac{1}{8\pi^2}\left(2C_2g^2 -\bar\lambda^2\right) \\
    &+\frac{2}{(16\pi^2)^2}\left[-\bar\lambda^4-\frac{N_f}{2} N\bar\lambda_1^4\right.\\
    &\left.+2g^4(C_2S_N(R)+2C_2^2-3C_N(G)C_2)
    \right],
    \label{eq:gamma_Q_m_2}
    \end{split}
\end{equation}
for $Q_a, \bar Q_a$ $(a=1,2,3)$, $Q_\alpha, \bar Q_\alpha$ $(\alpha=4, \cdots , N_f/2)$
and $Q_k, \bar Q_k$ $(k=N_f/2+1, \cdots , N_f)$, and
\begin{equation}
\begin{split}
    \gamma^2_\Phi&=\frac{1}{8\pi^2}\left(3N\lambda^2_1+\left(N_f/2-3\right)N\lambda_2^2\right)\\
    &+\frac{2}{(16\pi^2)^2}\left[-6N\lambda_1^4-2\lambda_2^4 N(N_f/2-3) N\right.\\
    &\left. +g^2\lambda_1^2C_212N+g_c^2\lambda_1^2C_2'12+g^2\lambda_2^2 4N(N_f/2-3)C_2
    \right]
     , \\[1ex]
    \gamma^2_{\bar\Phi}&=\frac{1}{8\pi^2}\bar\lambda^2N N_f/2\\
    &+\frac{2}{(16\pi^2)^2}\left[-2\bar\lambda^4 N(N_f/2)+g^2\bar\lambda^2 4N(N_f/2)C_2
    \right]
     ,
\end{split}
\end{equation}
for $\Phi, \bar\Phi$, 
where $C_2'=4/3$, $C_N(G)=N$, $S_N(R)=N_f$, $C_3(G)=3$ and $S_3(R)=N+6$.

\bibliography{bib}
\bibliographystyle{utphys}

\end{document}